\documentclass[conference]{IEEEtran}

\usepackage{graphicx}
\usepackage{algorithm}
\usepackage{algpseudocode}
\usepackage{soul}
\usepackage{cite}
\usepackage{amsmath,amssymb,amsfonts}
\usepackage{graphicx}
\usepackage{textcomp}
\usepackage{xcolor}
\usepackage{enumitem}
\usepackage{array}
\usepackage{soul}
\usepackage{textcomp}
\usepackage{subcaption}
\usepackage{multirow}
\usepackage[font=small]{caption}
\usepackage{amsmath}
\usepackage{bm}
\usepackage{listings}
\usepackage{color}
\usepackage{xcolor}
\usepackage[symbol]{footmisc}
\usepackage{url}
\usepackage[hidelinks]{hyperref}
\usepackage{comment}
\usepackage{pifont}
\usepackage[table]{xcolor}
\usepackage{scrextend}
\deffootnote{0.5em}{0em}{\thefootnotemark}

\newcommand{\PreserveBackslash}[1]{\let\temp=\\#1\let\\=\temp}
\newcolumntype{C}[1]{>{\PreserveBackslash\centering}p{#1}}
\newcolumntype{R}[1]{>{\PreserveBackslash\raggedleft}p{#1}}
\newcolumntype{L}[1]{>{\PreserveBackslash\raggedright}p{#1}}

\newcommand{\xmark}{\ding{55}}%

\lstdefinestyle{mystyle}{
    backgroundcolor=\color{backcolour},   
    commentstyle=\color{codegreen},
    keywordstyle=\color{magenta},
    numberstyle=\tiny\color{codegray},
    stringstyle=\color{codepurple},
    basicstyle=\ttfamily\footnotesize,
    breakatwhitespace=false,         
    breaklines=true,                 
    captionpos=b,                    
    keepspaces=true,                 
    numbers=left,                    
    numbersep=5pt,                  
    showspaces=false,                
    showstringspaces=false,
    showtabs=false,                  
    tabsize=2
}

\definecolor{dkgreen}{rgb}{0,0.6,0}
\definecolor{gray}{rgb}{0.5,0.5,0.5}
\definecolor{mauve}{rgb}{0.58,0,0.82}
\definecolor{gray}{rgb}{0.4,0.4,0.4}
\definecolor{darkblue}{rgb}{0.0,0.0,0.6}
\definecolor{lightblue}{rgb}{0.0,0.0,0.9}
\definecolor{cyan}{rgb}{0.0,0.6,0.6}
\definecolor{darkred}{rgb}{0.6,0.0,0.0}
\definecolor{codegreen}{rgb}{0,0.6,0}
\definecolor{codegray}{rgb}{0.5,0.5,0.5}
\definecolor{codepurple}{rgb}{0.58,0,0.82}
\definecolor{backcolour}{rgb}{0.95,0.95,0.92}

\lstdefinestyle{topofpage}{
  float=tp,
  floatplacement=tbp,
}

\begin{document}

\title{
Jack of All Scales: A Versatile FPGA Tensor Block for MXFP Precisions
\vspace{-0.4cm}
}

\author{
\IEEEauthorblockN{\small Marwan Mekhemer$^\dagger$, Ahmed Elsousy$^\dagger$, Balaji Venkatesh$^\dagger$, Raphael Rowley$^*$, Vaughn Betz$^*$, Nachiket Kapre$^\dagger$, Andrew Boutros$^\dagger$}
$^\dagger$University of Waterloo ~~~~~~ $^*$University of Toronto\\
\{msmekhemer, aelsousy, b4venkat, nachiket, andrew.boutros\}@uwaterloo.ca, raphael.rowley@mail.utoronto.ca, vaughn@eecg.utoronto.ca
\vspace{-0.5cm}
}

\maketitle

\begin{abstract}
Modern deep learning workloads increasingly rely on narrow numerical formats to improve efficiency and reduce memory footprint. 
The recently standardized microscaling floating-point (MXFP) family of formats, including MXFP8, MXFP6, and MXFP4, offers a practical approach to low-precision inference, yet the digital signal processing (DSP) blocks in current FPGA architectures offer limited native support for these formats. 
In this work, we first present a comprehensive characterization of MXFP dot product implementations on Altera Agilex-5 FPGAs, exploring a range of strategies spanning pure soft logic, DSP blocks in fixed-point, floating-point, and tensor modes. 
Our results show that while the tensor mode delivers the highest arithmetic density for MXFP4 (E2M1) and MXFP6 (E2M3), it cannot implement MXFP6 (E3M2) or any MXFP8 precisions, forcing designers to fall back to lower-density alternatives. 
Motivated by this gap, we propose targeted modifications to the DSP block's internal tensor-mode architecture that enable native support for all MXFP precisions while retaining backward compatibility. 
We estimate the area cost of these modifications using a simplified version  of the Agilex-5 DSP block core implemented using the open-source ASAP7 PDK.
We evaluate a variety of modified DSP block designs that present a tradeoff between format coverage, arithmetic density, and area overhead.
Our preferred design point increases the DSP tile area by 36\%, corresponding to only 1.8\% of the total FPGA die area. We evaluate the device-level impact of our enhanced DSP block by comparing systolic array matrix multiplier implementations across all MXFP precisions, contrasting the best-available strategies on the existing architecture against designs leveraging our modified DSP block. 
Our results demonstrate an average throughput improvement of 4.2$\times$ across all supported MXFP formats.
\end{abstract}

\IEEEpeerreviewmaketitle
\fontsize{9.9pt}{11.2pt}\selectfont

\section{Introduction}
\label{sec:intro}
 
In recent years, deep learning (DL) models have been continuously growing in size and computational cost.
Large language models such as GPT-4~\cite{achiam2023gpt4}, Llama 3~\cite{grattafiori2024llama}, and DeepSeek-V3~\cite{liu2024deepseekv3} contain hundreds of billions of parameters, placing significant demands on memory bandwidth and compute throughput.
Reducing the numerical precision of arithmetic operations is one of the most effective ways to alleviate these bottlenecks;
narrower formats reduce memory bandwidth requirements and decrease the silicon area of multiply-accumulate units, enabling higher peak throughput per unit area.
While 8-bit integer (\texttt{int8}) quantization has become widely adopted for inference~\cite{wu2020integer}, recent studies have shown that narrow floating-point (\texttt{fp}) formats can achieve comparable inference accuracy at 6 or even 4 bits per element~\cite{mxfp2023paper}.
Commercial architectures have also started to adopt these formats. 
For example, Nvidia's Blackwell architecture natively supports \texttt{fp4} and \texttt{fp6} operations in its fifth-generation tensor cores~\cite{blackwell2024}. 
To standardize narrow floating-point formats across the industry, an alliance including AMD, Arm, Intel, Meta, Microsoft, Nvidia, and Qualcomm defined the Microscaling (MX) specification under the Open Compute Project (OCP)~\cite{ocpmxfpspec2023}.
This specification defines a family of block-scaled floating-point formats, including MXFP8, MXFP6, and MXFP4, that provide a practical and interoperable approach to low-precision DL.
 
The reconfigurable fabric of field-programmable gate arrays (FPGAs) makes them particularly suited to a landscape with many competing and evolving numerical standards.
The narrow bit widths of MXFP formats make them amenable to soft-logic implementations using lookup tables (LUTs), and indeed this approach can be used to support any MXFP precision on any FPGA.
On the other hand, digital signal processing (DSP) blocks can deliver higher arithmetic density (i.e., more operations per unit area) for these operations, especially with the DL-targeted tensor compute modes introduced in recent FPGAs such as Altera's Stratix-10 NX~\cite{langhammer2021stratix} and Agilex-5~\cite{agilex5dsp}.
Mapping MXFP arithmetic to DSP blocks also frees the soft-logic fabric to implement other system and accelerator components, improving overall design efficiency.
 
However, current FPGA DSP blocks were not designed with MXFP support in mind.
While these blocks can still be repurposed to implement portions of MXFP arithmetic datapaths, as we demonstrate in this work, there are opportunities to significantly improve their arithmetic density through targeted changes to the internal DSP block architecture.
For instance, the tensor mode of the Altera Agilex-5 DSP block, its highest-density operating mode, can implement MXFP4 (E2M1) and MXFP6 (E2M3) dot products, but cannot be used for MXFP6 (E3M2) or any MXFP8 precision.
Designers targeting these formats must fall back to lower-density DSP modes or soft logic, incurring substantial area and throughput penalties.
 
This work presents a comprehensive study of MXFP support on state-of-the-art FPGAs.
We first characterize a wide range of MXFP dot product implementations on Altera Agilex-5 FPGAs, exploring strategies that use soft logic and DSP blocks in fixed-point, floating-point, and the high-density tensor modes, to identify the best implementation approach for each MXFP precision.
We then propose architectural changes to the DSP block's tensor mode that enable native support for MXFP precisions, including MXFP8, MXFP6, and MXFP4, while retaining backward compatibility with existing operating modes.
We estimate the area cost of these modifications using a simplified Agilex-5-like DSP block core implemented using 7nm standard cells, with full-custom programmable routing interfaces optimized using COFFE~\cite{COFFE2_2019}.
Finally, we evaluate the device-level impact of our enhanced DSP block by comparing systolic array matrix multiplier implementations targeting the baseline and modified DSP blocks.
In summary, our contributions are:
\begin{itemize}[leftmargin=*, itemindent=0pt, labelsep=3pt, align=left]
    \item A detailed characterization of MXFP dot product units mapped to the Altera Agilex-5 fabric, identifying the most efficient implementation style for each MXFP precision.
    \item DSP block tensor mode architectural enhancements for native MXFP support.
    \item A case study demonstrating the gains of the modified DSP block for systolic arrays using different MXFP precisions.
\end{itemize}
All our MXFP dot product implementations, DSP block designs, and systolic array benchmarks presented in this work are open-sourced to facilitate future research in this direction\footnote{Repository link: \href{https://github.com/boutros-lab/fpga-mxfp}{\texttt{https://github.com/boutros-lab/fpga-mxfp}}}.

\section{Background \& Related Work}
\label{sec:background}

\subsection{MXFP Formats}
\label{sec:mxfp_formats}
 
A floating-point number is represented by three fields: a sign bit $S$, an $e$-bit exponent field $E$, and an $m$-bit mantissa field $M$.
The value of the number is determined by whether the exponent is zero (subnormal) or nonzero (normal) as follows:
\begin{equation*}
(-1)^S \times 2^{E - \text{bias}} \times (1 + 2^{-m} \times M) ~~~~~~~ \text{if } E > 0 \text{ (normal)}
\end{equation*}
\vspace{-0.5cm}
\begin{equation*}
(-1)^S \times 2^{1 - \text{bias}} \times (2^{-m} \times M) ~~~~~~~~~ \text{if } E = 0 \text{ (subnormal)}
\end{equation*}
where $\text{bias} = 2^{e-1} - 1$.
Throughout this paper, we use the notation E$e$M$m$ to denote a floating-point format with $e$ exponent bits and $m$ mantissa bits (e.g., E4M3 denotes 4 exponent bits and 3 mantissa bits in addition to a sign bit).

The MX specification~\cite{ocpmxfpspec2023} defines a family of block-scaled floating-point formats designed for efficient DL computation.
In an MX-compliant format, a block of $k$ elements share a single unsigned E8M0 scale factor $X$ (with a bias of 127), while each element $P_i$ is independently encoded as a narrow floating-point number.
The shared scale is a power-of-two value that captures the coarse dynamic range of the block, while each element encodes its own sign, exponent, and mantissa fields.

The MX specification defines several concrete formats. 
MXFP8 specifies two sub-formats: E5M2, which provides a wider dynamic range per element, and E4M3, which offers higher precision at the cost of reduced range.
E5M2 has special encodings for infinity and NaN, while E4M3 has a special encoding for NaN only.
MXFP6 specifies E3M2 and E2M3 sub-formats, and MXFP4 uses a single E2M1 format, with no special encodings for infinity or NaN.
In all cases, the block size $k$ is 32 and the shared E8M0 block scale compensates for the limited per-element dynamic range, providing an effective dynamic range comparable to \texttt{fp32}.
There are other similar \emph{micro} floating-point formats that adopt different block and shared scale organizations.
For example, Nvidia's NVFP4 uses an \texttt{fp8} E4M3 shared scale for a block of sixteen \texttt{fp4} numbers and applies another \texttt{fp32} scale per tensor~\cite{alvarez2025nvfp4}. 
Although we focus mainly on MXFP formats in this work, our proposed enhancements can be extended to support other block-scaled floating-point formats, which we leave out for future work.

The key operation defined by the MX specification is the dot product of two MX-compliant vectors.
Given two blocks $\mathbf{a}$ and $\mathbf{b}$, each with $k$ elements and respective shared scales $X_a$ and $X_b$, the dot product is computed as:
\begin{equation*}
    \mathbf{a} \cdot \mathbf{b} = 2^{(X_a + X_b - 254)} \sum_{i=1}^{k} P_{a,i} \times P_{b,i}
    \vspace{-0.15cm}
\end{equation*}
The elementwise $P_{a,i} \times P_{b,i}$ can be implemented by multiplying the narrow significands (i.e., mantissa + implicit one) and summing their exponents.
The per-element exponent sums are then used to align the significand products before summing them. 
The final result is converted to \texttt{fp32} format, and the shared block scales are applied by adding ($X_a+X_b-254$) to the \texttt{fp32} exponent with appropriate clipping logic.

\subsection{FPGA Implementations of Narrow FP Arithmetic}
\label{sec:related_work}
The idea of using block-scaled narrow formats to increase arithmetic density on FPGAs predates the MX specification.
Aydonat et al.~\cite{aydonat2017deeplearningaccelerator} introduced an FPGA-based deep learning accelerator (DLA) on Altera Arria-10 that used a 16-bit block floating-point (\texttt{bfp16}) representation in which a 5-bit exponent is shared across a block of $k$ 10-bit mantissas.
This format enabled denser packing of fixed-point multiplications in DSP blocks and achieved better inference accuracy compared to integer quantization.
The Microsoft Brainwave project~\cite{brainwave2019} further developed this idea by deploying a soft neural processing unit (NPU) on Altera Stratix-10 FPGAs that used Microsoft floating-point (MSFP) formats.
MSFP shares the same core principle as MXFP in which a group of elements share a common exponent, reducing the bit width of elementwise multiplications.
The Brainwave NPU achieved up to 35.9 tera floating-point operations per second (TFLOPS) on the largest monolithic Stratix-10 device using these formats, demonstrating the arithmetic density advantage of block-scaled representations in soft logic.
A follow-up study by Rouhani et al.~\cite{rouhani2020pushing} showed that ASIC implementations of MSFP formats achieve 2.8-16.4$\times$ higher multiply-accumulate density per unit area compared to 16-bit Brain floating-point (\texttt{bfloat16}) and 1.1-6.6$\times$ compared to \texttt{int8}.
With the standardization of the MX specification, several recent works have explored MXFP implementations on FPGAs.
Samson et al.~\cite{samson2024exploring} presented the first open-source FPGA implementation of MX-compliant arithmetic, evaluating the area-accuracy tradeoff across soft logic implementations of a wide variety of MX formats and block sizes on AMD/Xilinx UltraScale+ devices.
More recently, Abdurakhmanov and Fahmy~\cite{mxfpsysarr2025} explored soft logic systolic array architectures for MXFP6 and MXFP8 formats on AMD UltraScale+ FPGAs, comparing exact accumulation (using a Kulisch-style wide fixed-point accumulator) and \texttt{bfloat16} accumulation, and achieving up to 0.57 and 0.5 TFLOPS for MXFP6 and MXFP8, respectively.
In contrast to these prior studies, our work characterizes different approaches for implementing MXFP arithmetic using soft logic and/or DSP blocks in different modes, including the DL-targeted tensor modes in state-of-the-art Altera FPGAs. 

\subsection{Prior Work on DSP Block Architecture Enhancements}
\label{sec:dsp_related}
 
Several prior works have investigated modifications to FPGA DSP block architectures to improve their efficiency for DL workloads.
The authors of~\cite{boutros2018embracing} were the first to propose DSP microarchitecture optimizations for low-precision DL, presenting an enhanced DSP block based on the Altera Arria-10 architecture that can pack 2$\times$ as many 9-bit and 4$\times$ as many 4-bit multiplications compared to the baseline block at a cost of 12\% increase in block area (0.6\% total FPGA die area increase).
PIR-DSP~\cite{rasoulinezhad2019pir} proposed a set of modifications to the Xilinx DSP48E2 block that introduce a runtime decomposable multiplier supporting multiple precisions and enhanced inter-DSP chaining for systolic structures.
Arora et al.~\cite{arora2022tensor} proposed Tensor Slices, which replace a portion of the FPGA's programmable logic with larger hard blocks containing systolic arrays of processing elements that support multiple tensor operations and different precisions including \texttt{int8}, \texttt{int16}, \texttt{fp16}, and \texttt{bfloat16}.
The Altera Stratix-10 NX~\cite{langhammer2021stratix} was the first commercial FPGA to adopt DL-optimized tensor blocks, replacing the conventional DSP block modes with high-density \texttt{int8} and \texttt{int4} dot product units, and the Agilex-5 DSP block subsequently integrated a similar tensor mode alongside the conventional DSP modes.
All of these prior efforts focused on integer and standard floating-point precisions.
To the best of our knowledge, no prior work has investigated DSP block architecture modifications to natively support MXFP formats, which is the focus of this paper.

\subsection{Agilex-5 Variable Precision DSP Block}
\label{sec:agilex5_dsp}

\begin{figure}
    \centering{\includegraphics[width=\columnwidth]{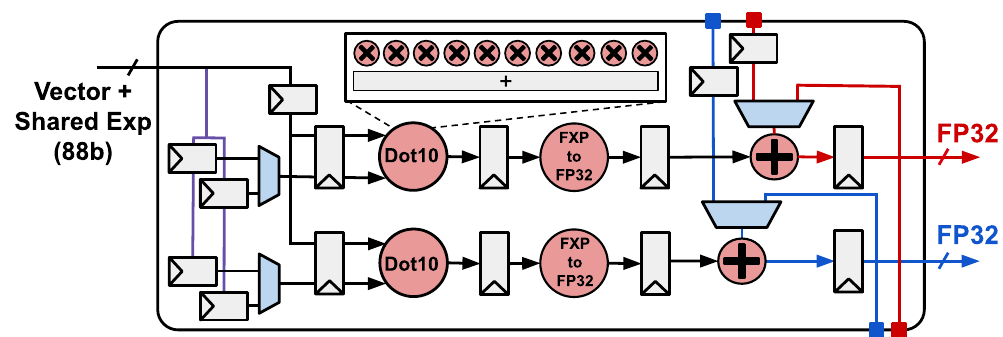}}
    \caption{Details of the Agilex-5 DSP floating-point tensor mode.} 
    \vspace{-0.4cm}
    \label{fig:baseline_aitb}
\end{figure}

This work focuses on the DSP block in the Altera Agilex-5 device.
This DSP block supports a variety of fixed-point, floating-point, and DL-targeted tensor modes of operation.
In the fixed-point mode, it can be configured to implement two independent 18$\times$18 bit, sum-of-two 18$\times$18, one 27$\times$27, or sum-of-six 9$\times$9 multiplications.  
In the floating-point mode, the block supports \texttt{fp32} and \texttt{fp16} arithmetic, including multiply, multiply-add, and multiply-accumulate operations.
In the \texttt{fp16} multiply-add mode, the block performs two \texttt{fp16} multiplications and sums their results into an \texttt{fp32} output.

The tensor mode, illustrated in Fig.~\ref{fig:baseline_aitb}, provides the highest arithmetic density.
It implements two 10-lane dot products ($\mathbf{a} \cdot \mathbf{b}$ and $\mathbf{a} \cdot \mathbf{c}$) that share a broadcast input vector operand~($\mathbf{a}$).
The other two operands~($\mathbf{b}$ and $\mathbf{c}$) are supplied from pre-loaded internal ping-pong register cascades.
The \emph{data feed} mechanism loads a cascade from the input data port by pausing computation for two cycles and amortizing this loading stall by reusing the loaded operands over many subsequent cycles.
Alternatively, the \emph{side feed} mechanism loads two elements at a time to one cascade over 10 cycles, while the other cascade is used for computation, without the need to stall computation.
This tensor mode has fixed-point and block floating-point variations.
The fixed-point tensor mode performs \texttt{int8} dot products and accumulates results in \texttt{int32}.
On the other hand, the block floating-point tensor mode uses an 8-bit shared exponent per 10-element vector operand and accumulates results in \texttt{fp32}.
Besides the clock and control signals, the DSP block in this mode uses a total of 104 input pins (80b input vector, 8b shared exponent, 16b side input feed), 64 output pins (two \texttt{fp32} results), and 64 dedicated accumulator chains between DSP blocks in the same column.
\section{MXFP Characterization Study}
\label{sec:characterization}

This section presents a comprehensive characterization of MXFP dot product implementations on Agilex-5, covering all five MXFP sub-formats (E2M1, E2M3, E3M2, E4M3, and E5M2) across four implementation approaches: a pure soft-logic baseline and three strategies that utilize DSP blocks in different operating modes.
All designs implement a 32-element dot product ($k = 32$) with \texttt{fp32} output, and all implementation results are collected using Quartus Prime Pro 25.3 targeting the Agilex-5 A5EC065BB32AE4S device.

\subsection{Existing Baseline Implementations}
\label{sec:baseline}
 
As a baseline, we use the open-source MXFP dot product implementations from~\cite{samson2024exploring} and map them to the Agilex-5 device.
These designs are written in behavioral Verilog and are mapped entirely to soft logic by the CAD tools, without utilizing any DSP blocks.
We added \texttt{fp32} output conversion logic using FloPoCo~\cite{de2011flopoco} to these implementations to enable a fair comparison across all approaches.
 
We also noticed that the original implementations are unpipelined, resulting in operating frequencies ranging from 47 to 98 MHz on Agilex-5 after we added the \texttt{fp32} output conversion logic.
To ensure a fair comparison, we deeply pipelined these designs, increasing their operating frequencies to 516-713 MHz.
These optimized soft-logic baselines serve as the reference point against which we compare the DSP-based approaches described in the following subsections.

\begin{figure}[t]
    \centering{\includegraphics[width=0.78\columnwidth]{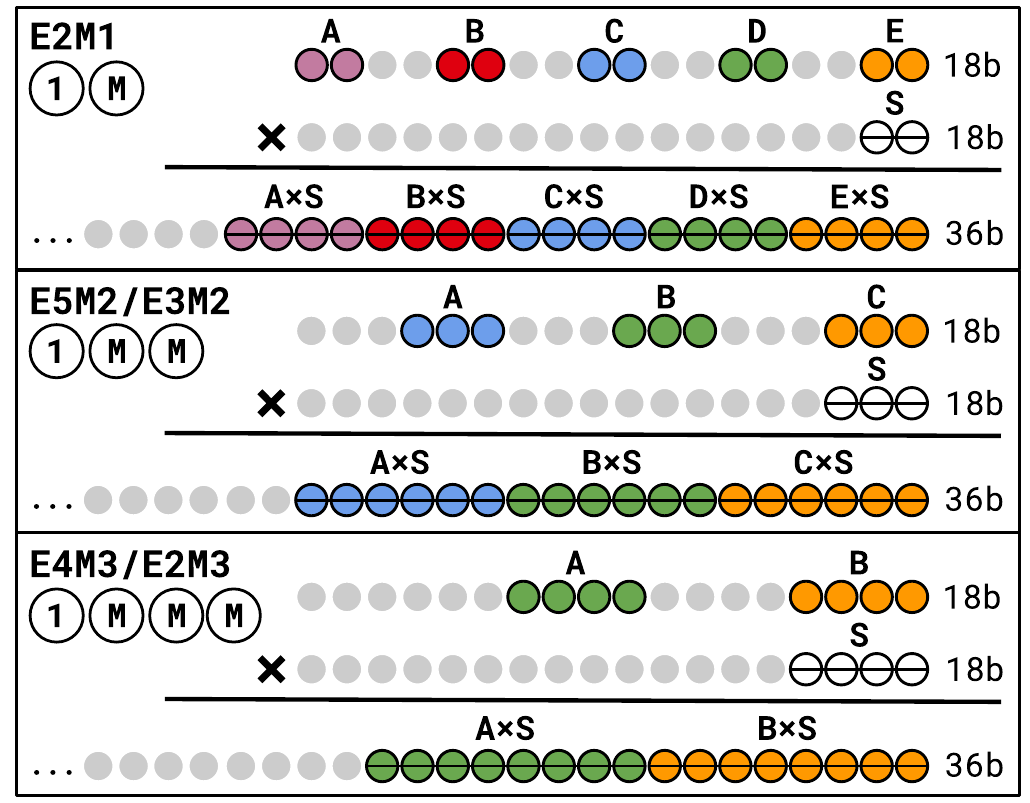}}
    \caption{Illustrations of multiplier packing with 2b, 3b, or 4b operands packed into an 18b input multiplied with a shared operand. The resulting products are extracted from the 36b multiplier output. An appropriately sized padding is introduced between packed operands such that their corresponding outputs do not overlap in the final result.} 
    \vspace{-0.4cm}
    \label{fig:packing}
\end{figure}

\begin{table*}[t]
\centering
\caption{Resource utilization results for different MXFP formats and implementation approaches on Agilex-5 FPGAs. The column labeled by $D$ represents the number of dot product units implemented per instance in the packed DSP and tensor mode approaches.}
\setlength{\tabcolsep}{3.5pt}
\begin{tabular}{|c|cccc|cccc|cccc|cccc|cccc|}
\hline
 & \multicolumn{4}{c|}{\textbf{Baseline}} & \multicolumn{4}{c|}{\textbf{Optimized Baseline}} & \multicolumn{4}{c|}{\textbf{Packed DSP}} & \multicolumn{4}{c|}{\textbf{\texttt{fp16} Vector Mode}} & \multicolumn{4}{c|}{\textbf{Tensor Mode}} \\
\textbf{Format} & \textbf{ALMs} & \textbf{DSPs} & \textbf{Freq.} & \textbf{$D$} & \textbf{ALMs} & \textbf{DSPs} & \textbf{Freq.} & \textbf{$D$} & \textbf{ALMs} & \textbf{DSPs} & \textbf{Freq.} & \textbf{$D$} & \textbf{ALMs} & \textbf{DSPs} & \textbf{Freq.} & \textbf{$D$} & \textbf{ALMs} & \textbf{DSPs} & \textbf{Freq.} & \textbf{$D$} \\
\hline
\textbf{E5M2} & 4,365 & 0 & 47 & 1 & 4,432 & 0 & 516 & 1 & 10,421 & 16 & 525 & 3 & 319 & 16 & 460 & 1 & -- & -- & -- & -- \\
\textbf{E4M3} & 3,100 & 0 & 52 & 1 & 3,287 & 0 & 539 & 1 & 4,841 & 16 & 581 & 2 & 660 & 16 & 455 & 1 & -- & -- & -- & -- \\
\textbf{E3M2} & 1,830 & 0 & 72 & 1 & 1,657 & 0 & 597 & 1 & 4,153 & 16 & 654 & 3 & 381 & 16 & 455 & 1 & -- & -- & -- & -- \\
\textbf{E2M3} & 1,688 & 0 & 81 & 1 & 1,579 & 0 & 658 & 1 & 1,962 & 16 & 850 & 2 & 368 & 16 & 446 & 1 & 420 & 4 & 458 & 2 \\
\textbf{E2M1} & 877 & 0 & 98 & 1 & 885 & 0 & 713 & 1 & 3,590 & 16 & 740 & 5 & 224 & 16 & 460 & 1 & 198 & 4 & 458 & 2 \\
\hline
\end{tabular}
\label{tab:char_results}
\vspace{-0.5cm}
\end{table*}

\subsection{Using Packed Fixed-Point Multipliers}
\label{sec:packed_mult}

In this approach, we exploit the fact that MXFP elementwise multiplications reduce to narrow unsigned fixed-point significand multiplications, which can be packed into the DSP block's fixed-point multiplier arrays.
Since the significand products must be aligned by their per-element exponent sums before summation, we cannot use the DSP block's sum-of-two 18$\times$18 or sum-of-six 9$\times$9 modes, which perform fixed summation of products without alignment.
Instead, we configure the DSP block in the two independent 18$\times$18 multiplication mode and pack multiple significand multiplications into each multiplier.
The DSP packing follows a similar approach to that presented in~\cite{langhammer2019extracting, xilinx2017int8}; multiple narrow multiplications are mapped to a single wide multiplier by spacing the operands such that their partial products do not overlap in the output word.
This requires that all multiplications packed into the same array share one operand, with the other operands concatenated and spaced appropriately across the input word as illustrated in Fig.~\ref{fig:packing}.
The number of significand multiplications that can be packed into a single 18$\times$18 array ($D$) depends on the significand bit width of the MXFP format.
Thus, narrower formats such as MXFP4 (E2M1) allow more multiplications per array than wider formats such as MXFP8 (E4M3).
 
After the packed multiplication results are extracted from the DSP output, the remaining operations are performed in soft logic.
Each significand product is converted to two's complement representation, shifted by the corresponding per-element exponent sum to align the products to a common fixed-point format, and then summed using a reduction tree.
The final result is converted to \texttt{fp32} and the shared block scales are applied.
Since each DSP block provides two multiplier arrays and each array produces $D$ results, 16 DSP blocks (in addition to soft logic resources) implement $D$ 32-element MXFP dot products.

\subsection{Using \texttt{fp16} Vector Mode}
\label{sec:fp16_vec}

In this approach, we convert all MXFP elements to \texttt{fp16} by zero-extending the mantissa and adjusting the exponent bias, then use the DSP block's native \texttt{fp16} arithmetic to perform the multiplications and additions.
Since \texttt{fp16} has a 5-bit exponent and 10-bit mantissa, it can represent any MXFP element value across all five sub-formats without loss of precision.
To implement a full 32-element MXFP dot product, we adopt the vector chaining technique of Langhammer and Pasca~\cite{arria10langhammer2015} that was originally proposed for \texttt{fp32} dot products on Arria-10 DSPs, and apply it to the \emph{vector one}, \emph{vector two}, and \emph{sum-of-two} \texttt{fp16} mode on Agilex-5~\cite{agilex5dsp}.
Each DSP block computes the sum of two \texttt{fp16} products, and multiple DSP blocks are chained into a recursive tree structure to sum partial results into a final \texttt{fp32} output, forming a fully-hardened dot product unit.
Minimal soft logic resources are used to adjust the input exponent biases during MXFP-to-\texttt{fp16} conversion and apply the shared block scales to the final \texttt{fp32} output.
This approach supports all MXFP sub-formats and minimizes soft logic utilization, but achieves lower arithmetic density than the packed fixed-point and tensor mode approaches since each DSP block computes only two multiplications regardless of the MXFP format used.

\subsection{Using Tensor Mode}
\label{sec:tensor_mode}

The floating-point tensor mode provides the highest arithmetic density of all DSP block modes, implementing two dot-10 operations for signed 8-bit operands with an 8-bit shared exponent per vector.
Since the internal dot product units sum the multiplication results without per-element alignment, the only way to use this mode for MXFP arithmetic is to first convert MXFP input vectors to signed fixed-point representation by aligning each element's significand according to its per-element exponent.
The shared exponent inputs of the tensor block are then used to internally apply the MXFP shared block scales.
 
This conversion to fixed-point is only feasible when the fixed-point values fit within the 8-bit signed DSP dot lanes.
For MXFP4 (E2M1) and MXFP6 (E2M3), the per-element exponent field is only 2 bits wide, resulting in shifts of at most $2^E-2 = 2$ positions (since $E=0$ indicates a subnormal value).
An additional bit is also needed for converting the aligned significand value from unsigned to two's complement format.
The aligned fixed-point values for these MXFP formats therefore fit within 8 bits (i.e., 5 bits for E2M1 and 7 bits for E2M3), making them compatible with the DSP block's tensor mode.
However, for MXFP6 (E3M2) and both MXFP8 sub-formats (E4M3 and E5M2), the wider per-element exponent range produces fixed-point values that are wider than 8 bits, preventing these formats from being mapped to the tensor mode.
 
DSP blocks operating in tensor mode can be cascaded using dedicated interconnect to accumulate the results of multiple dot-10 operations forming longer dot products.
A cascade of four DSP blocks implements two 32-element MXFP dot products that share one operand vector.
However, since $k = 32$ MXFP dot products still require $\lceil 32/10 \rceil = 4$ dot-10 units, the last DSP block has 8 out of 10 multipliers left unused, resulting in some underutilization.

\subsection{Characterization Results}
\label{sec:char_results}

\begin{figure}
    \vspace{-0.2cm}
    \centering{\includegraphics[width=0.8\columnwidth]{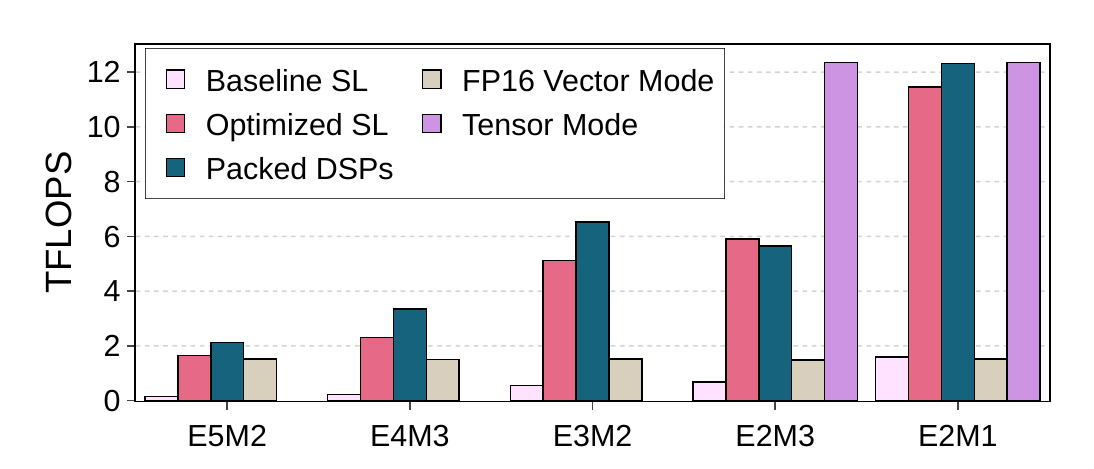}}
    \vspace{-0.2cm}
    \caption{Performance results for different MXFP formats and implementation styles when filling the Agilex-5 device with dot product units, assuming no frequency degradation as the device is filled up.}
    \label{fig:char_tflops}
    \vspace{-0.5cm}
\end{figure}

Table~\ref{tab:char_results} summarizes the resource utilization, operating frequency, and number of dot products per unit ($D$) for all MXFP formats across the four implementation approaches we discussed in this section.
Fig.~\ref{fig:char_tflops} presents the performance results in TFLOPS, computed by filling the entire Agilex-5 device with dot product units until the first resource type (ALMs or DSPs) is exhausted, and optimistically assuming that operating frequency does not degrade as utilization increases.
These performance results serve as an upper bound for achievable performance (i.e., \emph{peak performance}) that is more optimistic for implementations that use more soft logic resources (e.g., baseline and packed multipliers), which would typically suffer from a steeper frequency degradation as the device is filled up.
 
The optimized soft-logic baseline achieves 1.7-11.5 TFLOPS depending on the format, with narrower formats achieving higher throughput due to their smaller multiplier and adder tree area.
The packed fixed-point multiplier approach achieves the highest throughput for E5M2, E4M3, and E3M2 formats, reaching 2.1, 3.4, and 6.5 TFLOPS, respectively.
For these formats, it improves performance by 28-45\% compared to the optimized soft logic baseline while using 16-26\% fewer ALMs per dot product.
However, this approach still requires significant soft logic resources (1,384-3,474 ALMs per dot unit) for exponent alignment, two's complement conversion, and reduction trees.
 
The \texttt{fp16} vector mode provides consistent throughput of 1.5 TFLOPS across all five formats with minimal soft logic overhead (224-660 ALMs per dot product unit).
Since each DSP block computes only two \texttt{fp16} multiplications regardless of the MXFP element width, this approach does not benefit from narrower formats.
However, this DSP-based approach supports all five MXFP sub-formats while utilizing significantly less soft logic resources compared to the packed DSP approach.
 
The tensor mode achieves the highest throughput for E2M3 and E2M1 formats, both reaching 12.4 TFLOPS.
These results are 2.1$\times$ and 1.1$\times$ higher than the optimized soft logic baseline for E2M3 and E2M1, respectively.
Realistically, these performance gaps would be significantly wider as the soft-logic-based approaches are expected to experience a greater operating frequency degradation as the device is filled with dot product units. 
However, as discussed in Section~\ref{sec:tensor_mode}, this mode is limited to MXFP4 and MXFP6 (E2M3) because wider per-element exponent ranges produce fixed-point values that exceed the 8-bit multiplication lanes in the DSP block.
For MXFP6 (E3M2) and both MXFP8 formats, designers must fall back to the packed fixed-point mode, which delivers significantly lower throughput.
This limitation motivates the DSP block architecture modifications we propose to the tensor mode of operation in Section~\ref{sec:aitb-changes}.

\section{DSP Block Architecture Modifications}
\label{sec:aitb-changes}

This section describes our proposed modifications to the Agilex-5 DSP block's tensor mode to enable native support for MXFP precisions.
We present our design iterations, starting from a straightforward extension of the existing fixed-point input approach and progressively refine the architecture to balance format coverage, arithmetic density, and area overhead.

\subsection{Baseline Implementation and Methodology}
\label{sec:baseline_block}

We implement a simplified version of the Agilex-5 DSP block that supports the baseline floating-point tensor mode (see Fig.~\ref{fig:baseline_aitb}).
This block consists of two dot-10 units, ping-pong data reuse register banks, fixed-point to \texttt{fp32} conversion, an \texttt{fp32} adder/accumulator (generated using FloPoCo~\cite{de2011flopoco}), and shared exponent handling logic.
This simplified version does not implement the other fixed-point and floating-point modes of the Agilex-5 DSP block.
Implementing a full-fidelity model of the commercial DSP block, including all operating modes, is beyond the scope of this work and is not necessary since our proposed modifications target only the tensor mode datapath.
Other modes in the commercial block reuse the same multiplier arrays and add SRAM-controlled reconfiguration circuitry for mode selection.
Since the other modes remain unchanged in our proposed modifications, all area overheads reported in this section are overestimates; when considering the additional circuitry for all supported modes, our proposed modifications would represent a smaller fraction of the total block footprint.

We implement the DSP block core in standard cell using the ASAP 7nm process design kit (PDK)~\cite{vashishtha2017asap7}.
We run the ASIC implementation tools through the HAMMER flow~\cite{liew2022hammer}, which we configure to use Cadence Genus 21.17 for synthesis and Cadence Innovus 21.17 for place and route (PnR).
Since the Agilex-5 DSP block in the floating-point tensor mode operates at 458~MHz (see Table~\ref{tab:char_results}), we set a target clock period of 2 ns.
We first run synthesis with an unconstrained floorplan area to obtain the total cell area, then re-run synthesis and PnR with a floorplan area constrained to 1.5$\times$ the total cell area to obtain final timing and area results.
The ASAP7 PDK cells are upsized by 4$\times$ in each dimension to bypass academic license constraints that do not allow below 20nm features~\cite{vashishtha2017asap7}, so all reported area results are scaled back to get 7nm area estimates.
This methodology is applied consistently to all design variations we explore.

\begin{table}[t]
    \centering
    \caption{Area breakdown of our simplified baseline DSP block implementing the Agilex-5 tensor mode of operation.}
    \label{tab:baseline_area}
    \begin{tabular}{lcc}
        \hline
        \textbf{Component} & \textbf{Area ($\mu m^2$)} & \textbf{Percentage}\\
        \hline
        Standard-cell Core &  1,262 & 69.7\%\\
        Switch Block &  216 & 11.9\%\\
        Connection Block &  134 & 7.4\%\\
        Local Crossbar &  191 & 10.5\%\\
        Dedicated Chain Buffers &  8 & 0.5\%\\
        \hline
        \textbf{Total} &  \textbf{1,811} & 100\%\\
        \hline
    \end{tabular}
    \vspace{-0.1cm}
\end{table}

In addition to the standard-cell core, the DSP tile includes interfaces to the programmable routing fabric, a local crossbar for DSP block inputs, buffers for dedicated chains between DSP blocks, and global routing circuitry (i.e., switch block and connection block).
We use the COFFE tool~\cite{COFFE2_2019} to optimize the transistor sizing of these full-custom components assuming 25\% populated local crossbar, 104 inputs, 64 outputs, and 64 dedicated chains between DSPs.
Table~\ref{tab:baseline_area} shows the area breakdown of the baseline DSP block tile, combining the standard-cell core area from PnR with the full-custom routing and interface circuitry area from COFFE.

The design modifications presented in the following subsections are guided by two main goals: (1) preserving or reducing the number of DSP block input/output pins to avoid any additional area overhead from added routing interfaces, and (2) having no negative impact on tensor mode operating frequency.

\begin{table}[t]
    \centering
    \caption{Significand multiplier size and aligned product width for each MXFP format for the native MXFP inputs approach. The aligned product width determines the CSA reduction tree width of the dot units in the DSP block.}
    \label{tab:mxfp_widths}
    \begin{tabular}{lccccc}
        \hline
        & \textbf{MXFP4} & \multicolumn{2}{c}{\textbf{MXFP6}} & \multicolumn{2}{c}{\textbf{MXFP8}} \\
        & \textbf{E2M1} & \textbf{E2M3} & \textbf{E3M2} & \textbf{E4M3} & \textbf{E5M2} \\
        \hline
        \textbf{Significand Mult.} & 2$\times$2 & 4$\times$4 & 3$\times$3 & 4$\times$4 & 3$\times$3 \\
        \textbf{Aligned Prod. (bits)} & 9 & 13 & 19 & 37 & 67 \\
        \hline
    \end{tabular}
    \vspace{-0.25cm}
\end{table}

\begin{table*}[t]
    \centering
    \caption{Area results for each design iteration. Dot sizes are per dot unit (two per DSP block). Interface area includes switch block, connection block, local crossbar, and dedicated chain buffers from COFFE. The \xmark~symbol means this format is not supported.}
    \label{tab:mod_results}
    \begin{tabular}{L{3.8cm} C{0.5cm}C{0.5cm}C{0.5cm}C{0.5cm}C{0.5cm}C{0.5cm} C{1.8cm}cc c c}
        \hline
        & \multicolumn{6}{c}{\textbf{Dot Size per Format}} & \textbf{Core} & \textbf{Interface} & \textbf{Total Tile} & \textbf{Area} & \textbf{Freq} \\
        \textbf{Design Iteration} & \textbf{INT8} & \textbf{E2M1} & \textbf{E2M3} & \textbf{E3M2} & \textbf{E4M3} & \textbf{E5M2} & \textbf{Area ($\mu m^2$)} & \textbf{Area ($\mu m^2$)} & \textbf{Area ($\mu m^2$)} & \textbf{Ratio} & \textbf{(MHz)}\\
        \hline
        \footnotesize{Baseline} & 10 & 10 & 10 & \xmark & \xmark & \xmark & 1,262 & 549 & 1,811 & 1.00$\times$ & 483\\
        (1) Fixed-point inputs & 10 & 16 & 11 & 8 & \xmark & \xmark & 1,773 & 549 & 2,322 & 1.28$\times$ & 490\\
        (2) MXFP inputs (all) & 10 & 16 & 12 & 12 & 8 & 8 & 2,448 & 549 & 2,997 &  1.65$\times$ & 475\\
        (3.1) MXFP inputs (no MXFP8) & 10 & 16 & 12 & 12 & \xmark & \xmark & 1,825 & 549 & 2,374 & 1.31$\times$ & 482\\
        (3.2) MXFP inputs (no E5M2) & 10 & 16 & 12 & 12 & 8 & \xmark & 2,069 & 549 & 2,618 & 1.45$\times$ & 486\\
        (3.3) MXFP inputs (4$\times$E5M2) & 10 & 16 & 12 & 12 & 8 & 4 & 2,336 & 549 & 2,885 & 1.59$\times$ & 484\\
        (4) MXFP inputs (no E5M2) & 8 & 16 & 12 & 12 & 8 & \xmark & 1,958 & 509 & 2,467 & 1.36$\times$ & 482\\
        \hline
    \end{tabular}
    \vspace{-0.4cm}
\end{table*}

\subsection{Iteration 1: Extending the Fixed-Point Input Approach}
\label{sec:iter1}
 
We start from the observation that the baseline DSP tensor mode implements E2M1 and E2M3 by converting MXFP input vectors to signed fixed-point representation in soft logic before feeding them into the DSP block's 8-bit dot product lanes.
The signed fixed-point representations are 5 bits for E2M1 elements and 7 bits for E2M3 elements. 
This means that for dot-10 operations, an input vector is using only 50 and 70 out of the 80 data input pins for E2M1 and E2M3, respectively.
Therefore, there is an opportunity to use the remaining input pins to provide up to 16-element E2M1 or 11-element E2M3 vectors as inputs to the DSP block.
Thus, we add five small 5$\times$5 multipliers and one 7$\times$7 multiplier to each of the two dot units in the DSP block, enhancing the arithmetic density to 2$\times$dot-16 for E2M1 and 2$\times$dot-11 for E2M3 per block.
This reduces the number of DSP blocks required for a $k = 32$ MXFP dot product from 4 (in the baseline) to 2 for E2M1 and 3 for E2M3.

\begin{figure}
    \centering{\includegraphics[width=\columnwidth]{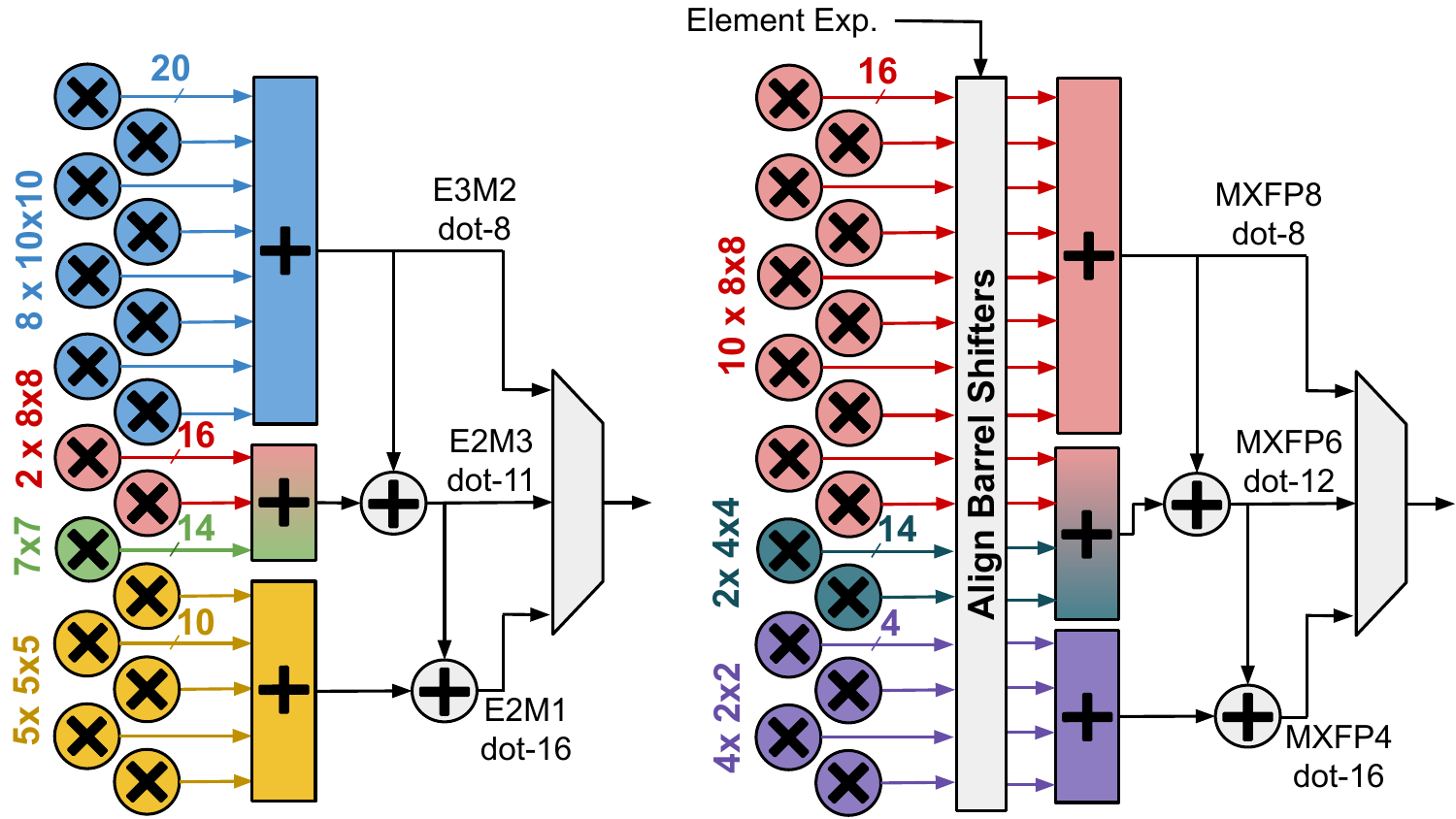}}
    \caption{Left: Modified dot product unit in our proposed DSP block with fixed-point inputs (iteration 1).
            Right: Modified dot product unit in our DSP block with native MXFP inputs (iteration 2).} 
    \vspace{-0.4cm}
    \label{fig:iteration1}
\end{figure}

For MXFP6 (E3M2), the signed fixed-point representation is 10 bits wide.
We add support for this format by expanding eight of the ten existing multipliers per dot unit from 8$\times$8 to 10$\times$10, enabling 2$\times$dot-8 for E3M2 per block.
Expanding all ten multipliers to support dot-10 E3M2 would incur additional area overhead with no tangible benefit since a $k = 32$ dot product would still require $\lceil 32/10 \rceil = 4$ DSP blocks.
All these modes still require modest soft logic resources for MXFP-to-fixed-point input conversion.
Following the same logic for MXFP8, the fixed-point representations of E4M3 and E5M2 are 19 and 34 bits wide, respectively, which would require multipliers significantly larger than our expanded 10$\times$10, resulting in prohibitive area overheads for this approach.
In addition, there would be enough input pins to support only dot-4 E4M3 and dot-2 E5M2, which diminishes the arithmetic density advantage of the DSP tensor mode compared to other implementation approaches.
The left side of Fig.~\ref{fig:iteration1} illustrates the modified dot unit for this design iteration, which supports dot-16 E2M1, dot-11 E2M3, and dot-8 E3M2.
These modifications result in a 28\% increase in the DSP tile area as shown in Table~\ref{tab:mod_results}.

\subsection{Iteration 2: Native MXFP Inputs \& Narrow FP Multiplies}
\label{sec:iter2}
 
Another design approach is to modify the DSP block to accept operands in their native MXFP encoding and perform elementwise narrow floating-point multiplications instead.
This eliminates the external soft logic overhead for input conversion and results in an architecture that generalizes more naturally to support additional MXFP formats.
In this modified DSP, we multiply the unsigned significands using the existing integer multipliers.
Then, we add circuitry to convert the products to two's complement format and align them using barrel shifters controlled by the sum of per-element exponents.
The aligned products are then reduced using a carry-save adder (CSA) tree, converted to \texttt{fp32}, and the shared block scales are applied.

The CSA reduction tree in the baseline block sums ten 16-bit products.
However, for our modified block, the CSA reduction tree needs to be expanded to sum up $k'$ aligned products, where $k'$ is the length of the dot product implemented per DSP block.
Table~\ref{tab:mxfp_widths} summarizes the sizes of the significand multiplier needed and aligned product bit width (i.e., bit width of CSA reduction tree inputs) for each of the MXFP formats.
This shows that the MXFP8 formats (E4M3 and E5M2) require a significant expansion in the CSA reduction tree to sum up 37-bit and 67-bit operands compared to 16-bit operands in the baseline block.

To satisfy the constraint of 80 data input pins, the DSP block can accept up to 16 MXFP4, 13 MXFP6, and 10 MXFP8 elements per input vector.
However, for a block size of $k = 32$, it is sufficient to have 16 MXFP4, 12 MXFP6, or 8 MXFP8 inputs to implement dot-32 in 2, 3, and 4 blocks, respectively.
This also minimizes the area overhead of added multipliers, two's complement converters, barrel shifters, and expanded CSA reduction tree inputs if they do not result in resource savings when implementing $k=32$ MXFP dot products.
To enable these dot lengths, besides the existing ten 8$\times$8 multipliers per dot unit, we introduce two additional 4$\times$4 multipliers (total of $10+2$ multipliers for dot-12 MXFP6) and four 2$\times$2 multipliers (total of $10+2+4$ multipliers for dot-16 MXFP4).
The right side of Fig.~\ref{fig:iteration1} illustrates the modified dot unit for this design iteration, which implements dot-16 E2M1, dot-12 E2M3/E3M2, and dot-8 E4M3/E5M2.
For the MXFP8 modes, we do not support special encodings for infinity/NaN.
As shown in Table~\ref{tab:mod_results}, these modifications result in a 65\% increase in DSP tile area.

\subsection{Iteration 3: MXFP8 Support Tradeoffs}
\label{sec:iter3}
 
MXFP8 formats introduce the widest barrel shifters and result in the largest increase in CSA reduction tree width, dominating the area overhead.
We therefore evaluate three design choices that present different tradeoffs between precision coverage, arithmetic density, and area overhead:
\begin{enumerate}[leftmargin=*, itemindent=0pt, labelsep=3pt, align=left]
    \item Drop MXFP8 support entirely, limiting the block to MXFP4 and MXFP6 formats.
    \item Support only the E4M3 variant of MXFP8 at dot-8 capacity. This is motivated by the observation that E4M3 achieves higher inference accuracy than E5M2 across many DL workloads~\cite{mxfp2023paper}, and inference is the primary use case for FPGAs in the DL domain~\cite{boutros2025field}. 
    \item Support E4M3 variant at dot-8 capacity and E5M2 at a reduced dot-4 capacity to reduce the area overhead of the large CSA reduction tree expansion (67-bit reduction operands).
\end{enumerate}

Table~\ref{tab:mod_results} shows that these design points result in a DSP tile that is 0.79$\times$, 0.87$\times$, and 0.96$\times$ the area of the one that supports all MXFP formats at highest capacity (from iteration 2).
We believe that the DSP block with support for only the E4M3 variant of MXFP8 offers a reasonable design tradeoff as it still supports the inference-targeted variant of MXFP8 and increases the DSP tile area by 45\% compared to the baseline.

\subsection{Iteration 4: Optimizing the Baseline Dot Length}
\label{sec:iter4}
 
We observe that most practical designs dot product implementations typically use a power-of-two number of lanes, while the baseline tensor mode implements dot-10.
For example, \texttt{int8} dot-16 and dot-32 require $\lceil 32/10 \rceil = 4$ and $\lceil 16/10 \rceil = 2$ DSP blocks with 8 and 4 of the 10 multipliers in the last block left unused, respectively.
By reducing the baseline dot length from dot-10 to dot-8, the last block is fully utilized and we can shrink two of the ten 8$\times$8 multipliers per dot to 4$\times$4 multipliers.
The smaller multipliers still support dot-12 for MXFP6 and dot-16 for MXFP4, preserving the density gains from earlier design iterations while clawing back some of the area overhead introduced by the wider MXFP format support.
After this change, the DSP block still has enough multiplier arrays that can be stitched together to realize other modes, such as the two 18$\times$18 and one 27$\times$27 fixed-point multiplication modes.
However, it is not fully backward compatible as it implements dot-8 instead of dot-10 operations in the baseline tensor mode.
In addition to savings in standard cell core area, this change also reduces the area of full-custom interfaces to the programmable routing since the number of input data pins is reduced from 80 (in the baseline) to 72 (for 12 MXFP6 elements).
This modification reduces the DSP tile area overhead to 36\% compared to the baseline (see Table~\ref{tab:mod_results}). 
We select this as our best design candidate for further evaluation.

With DSP tiles constituting around 5\% of the die area in DSP-rich devices\footnote{Based on publicly available estimates for Arria-10 devices~\cite{arria10langhammer2015}.}, this block-level area overhead translates to only 1.8\% increase in the entire FPGA fabric area.
This is an upper-bound for the area overhead introduced by our DSP block architecture modifications; 
for a commercial block implementing all other modes that we did not include in our baseline (and remain unchanged in our proposed block), the percentage increase in DSP tile area will be further diluted making this an even more appealing proposal.
All our DSP block designs can operate at above 458 MHz, which is the frequency of the Agilex-5 DSP block when configured in the floating-point tensor mode (see Table~\ref{tab:char_results}), and have the same latency in cycles.

Table~\ref{tab:mod_dot_results} shows the resource utilization of a $k=32$ MXFP dot module and the peak performance results\footnote{Assuming the frequency of a single instance is achievable for a full device.} of an Agilex-5 device comparing the best implementation approach using the baseline Agilex-5 DSP block (from Section~\ref{sec:characterization}) against tensor mode implementations using our modified DSP block with native MXFP support. 
The column labeled by $D$ represents the number of dot product operations implemented per module instance. 
The gray cells show the results of our DSP block variation that supports dot-8 E5M2 (iteration 2 in Table~\ref{tab:mod_results}) to demonstrate its potential gains in case E5M2 support is desired at the cost of additional area overhead.
These results show that our DSP block modifications enhance the device peak performance by 5.9$\times$, 3.6$\times$, and 2.5$\times$ for E5M2, E4M3, and E3M2 formats, respectively, while using minimal soft logic resources for staggering inputs of subsequent blocks in a DSP cascade.
For the E2M3 and E2M1 formats that can be mapped to the tensor mode of the baseline DSP block, our proposed modifications still enhance their arithmetic density (dot-16 and dot-12 instead of dot-10), achieving 1.3$\times$ and 2$\times$ higher TFLOPS, respectively. 

\begin{table}[t]
\centering
\caption{Resource utilization and device peak performance comparison for an Agilex-5 FPGA with the baseline DSP and our proposed DSP block.}
\setlength{\tabcolsep}{3.5pt}
\begin{tabular}{|c|cccc|cccc|} 
\hline
 & \multicolumn{4}{c|}{\textbf{Best using Baseline DSP}} & \multicolumn{4}{c|}{\textbf{Using our Proposed DSP}} \\
\textbf{Format} & \textbf{ALMs} & \textbf{DSPs} & \textbf{$D$} & \textbf{TFLOPS} & \textbf{ALMs} & \textbf{DSPs} & \textbf{$D$} & \textbf{TFLOPS} \\
\hline
\textbf{E5M2} & 10,421 & 16 & 3 & 2.1 & \cellcolor{gray!20}210 & \cellcolor{gray!20}4 & \cellcolor{gray!20}2 & \cellcolor{gray!20}12.4\\
\textbf{E4M3} & 4,841 & 16 & 2 & 3.4 & 210 & 4 & 2 & 12.4\\
\textbf{E3M2} & 4,153 & 16 & 3 & 6.5 & 96 & 3 & 2 & 16.5\\
\textbf{E2M3} & 420 & 4 & 2 & 12.4 & 96 & 3 & 2 & 16.5\\
\textbf{E2M1} & 198 & 4 & 2 & 12.4 & 39 & 2 & 2 & 24.8\\
\hline
\end{tabular}
\label{tab:mod_dot_results}
\vspace{-0.5cm}
\end{table} 
\section{Case Study: MXFP Systolic Arrays}
\label{sec:case-study}

To evaluate the device-level impact of our proposed DSP block modifications, we implement a systolic array for matrix multiplication on Agilex-5.
Matrix multiplication is the dominant computational kernel in DL workloads, the primary use case for MXFP formats.
We compare systolic array designs using the baseline DSP block architecture against those using our modified DSP block to quantify the throughput gains across different MXFP formats.

\subsection{Systolic Array Architecture}
\label{sec:systolic_arch}

\begin{figure}
    \centering{\includegraphics[width=0.8\columnwidth]{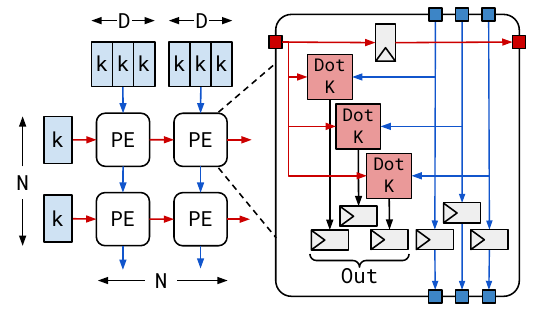}}
    \caption{Illustration of the systolic array and processing element. } 
    \vspace{-0.4cm}
    \label{fig:sysarr}
\end{figure}
 
Our systolic array, illustrated in Fig.~\ref{fig:sysarr}, multiplies an $N \times k$ matrix by a $k \times (N \times D)$ matrix, where $k = 32$ is the MXFP block size and $D$ is the number of 32-element MXFP dot product operations computed per processing element (PE).
We sweep the value of $N$ to implement progressively larger systolic arrays until the device resources are exhausted or the design cannot route.
 
For the baseline Agilex-5 DSP block, we select the best implementation approach for each MXFP format based on the characterization results from Section~\ref{sec:char_results}: the tensor mode for MXFP4 (E2M1) and MXFP6 (E2M3), and the packed fixed-point multiplier approach for all other formats.
In the packed DSP configuration, vectors from both matrix operands are streamed into the systolic array at the same time.
In the tensor mode configuration, vectors from one matrix are loaded into the DSP block's internal register banks while vectors from the other matrix are streamed into the array.

\subsection{Verification and Implementation Methodology}
\label{sec:systolic_method}
 
To verify the functional correctness of the systolic array designs using our modified DSP block, we use the block's RTL implementation as a simulation model.
This is the same RTL that we push through the ASIC implementation flow for area estimation in Section~\ref{sec:aitb-changes}.
For the systolic array implementation in Quartus, we exploit the fact that our modifications do not increase the number of input/output pins of the DSP block or its maximum operating frequency.
We instantiate DSP block IPs and connect their ports according to the modified tensor mode interface, effectively using the Agilex-5 DSP block as a pin-compatible stand-in for our modified block.
This allows Quartus to perform placement and routing of the full systolic array design, capturing the effect of the new block on the overall design's resource utilization, routability, and operating frequency, while the internal DSP block behavior is verified separately through RTL simulation.

\subsection{Case Study Results}
\label{sec:systolic_results}

\begin{figure}
    \centering{\includegraphics[width=\columnwidth]{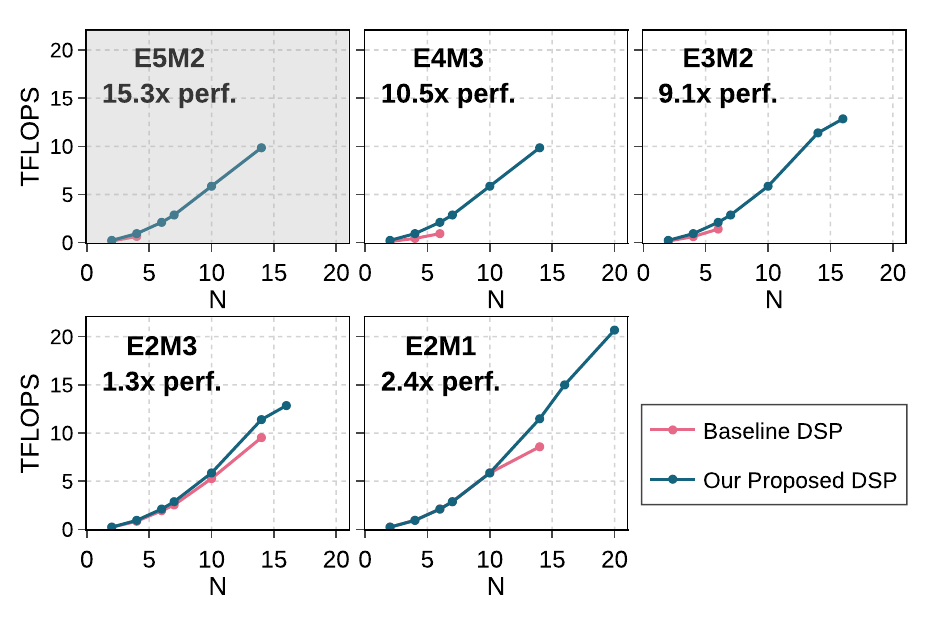}}
    \vspace{-0.5cm}
    \caption{Performance results for the systolic array design targeting the baseline Agilex-5 DSP block and our proposed DSP block. E5M2 results are using the modified DSP block from design iteration 3.3 (see Table~\ref{tab:mod_results}) to highlight its potential performance gains at a higher DSP block silicon area footprint.} 
    \vspace{-0.4cm}
    \label{fig:sysarr_results}
\end{figure}

Fig.~\ref{fig:sysarr_results} shows the throughput in TFLOPS as a function of the systolic array size $N$ for each MXFP format, comparing designs using the baseline Agilex-5 DSP block against those using our proposed modified DSP block.
We include E5M2 results using the DSP block variant that supports this format (Iteration 2), which has a larger area overhead than our recommended design (Iteration 4) to illustrate the potential gains if full MXFP8 coverage is desired.

The largest gains are observed for the MXFP8 and MXFP6 (E3M2) formats, which cannot use the tensor mode on the baseline DSP block and must fall back to the packed fixed-point multiplier approach.
Our modified DSP block enables the tensor mode for these formats, resulting in 15.3$\times$, 10.5$\times$, and 9.1$\times$ higher throughput for E5M2, E4M3, and E3M2, respectively, at the largest systolic array size that fits on the device.
These gains stem from two factors: the higher arithmetic density of the tensor mode compared to the packed multipliers approach, and the significantly lower soft logic utilization per PE, which allows larger systolic arrays to be instantiated.

For E2M3 and E2M1, which use the tensor mode in the baseline DSP block, the gains are more modest at 1.3$\times$ and 2.4$\times$, respectively.
These improvements come from the increased dot length per DSP (dot-12 vs. dot-10 for E2M3 and dot-16 vs. dot-10 for E2M1), which reduces the number of DSP blocks required per PE and allows more PEs to fit on the device.
The performance gain for E2M1 is greater than E2M3 because the proposed modifications reduce the number of required DSP blocks for $k = 32$ from 4 in the baseline DSP, to 2 and 3 for E2M1 and E2M3, respectively.

The largest systolic array designs across all MXFP formats use more than 90\% of the device's DSP blocks and can still run at around 400 MHz while using only 22-32\% of the fabric's soft logic resources.
This highlights that significant soft logic resources remain available for implementing other accelerator components such as control logic, data marshalling, and memory interfaces.
Across all supported MXFP formats (MXFP4, MXFP6, and E4M3 MXFP8), our proposed DSP block enhances throughput by 4.2$\times$ on average.

\section{Conclusion}
\label{sec:conc}

Microscaling floating-point formats are emerging as a practical approach to low-precision deep learning, yet current FPGA DSP blocks offer limited native support for these formats.
In this work, we presented a comprehensive study of MXFP support on state-of-the-art FPGAs, spanning characterization of existing implementation approaches to the proposal and evaluation of targeted DSP block enhancements.
 
Our characterization of MXFP dot product implementations revealed that the DSP block's tensor mode delivers the highest arithmetic density for MXFP4 (E2M1) and MXFP6 (E2M3), but cannot support MXFP6 (E3M2) or MXFP8 formats, forcing designers to fall back to lower-density alternatives.
To address this gap, we proposed modifications to the tensor mode architecture that enable native support for all MXFP6 and MXFP8 (E4M3) precisions while retaining backward compatibility.
Through an iterative design exploration, we identified a design point that achieves this expanded format coverage with a 36\% increase in DSP tile area, without degrading the operating frequency or changing the block's interface to the programmable routing.
Since DSP tiles occupy a small fraction of the total FPGA die, this translates to only a 1.8\% increase in overall die area.
Finally, we evaluated the impact of our modified DSP block on systolic array matrix multiplier implementations across all MXFP formats.
For formats that could not previously use the tensor mode (E5M2, E4M3, E3M2), our modified block enables throughput gains of 9.1-15.3$\times$ by replacing soft-logic-intensive packed DSP implementations with high-density tensor mode operations.
For formats already supported by the baseline tensor mode (E2M1, E2M3), the increased dot length per block yields gains of 1.3-2.4$\times$.
The complete set of RTL implementations, including all MXFP dot product units, modified DSP block designs, and systolic array benchmarks, are publicly available to support further exploration in this direction.

\section*{Acknowledgments}
The authors thank the Natural Sciences and Engineering Research Council of Canada (NSERC) for funding support.

\bibliographystyle{IEEEtran}
\bibliography{references}

\end{document}